# Anisotropic Hybridization Dynamics in the Quasi-One-Dimensional Kondo Lattice $CeCo_2Ga_8$ Revealed by Ultrafast Optical Spectroscopy

Ba-Lei Tan,[1] Chen Zhang,[1] Qi-Yi Wu,[1] Guo-Hao Dong,[2,3] Hao Liu,[1] Bo Chen,[1] Jiao-Jiao Song,[1] Xin-Yi Tian,[1] Ying Zhou,[1] Hai-Yun Liu,[4] Yu-Xia Duan,[1] You-Guo Shi,[2,3] and Jian-Qiao Meng[1,*]

[1] *School of Physics, Central South University, Changsha 410083, Hunan, China*
[2] *Beijing National Laboratory for Condensed Matter Physics, Institute of Physics, Chinese Academy of Sciences, Beijing 100190, China*
[3] *School of Physical Sciences, University of Chinese Academy of Sciences, Beijing 100049, China*
[4] *Beijing Academy of Quantum Information Sciences, Beijing 100085, China*
(Dated: Wednesday 12$^{th}$ March, 2025)

We investigate the ultrafast dynamics of the quasi-one-dimensional Kondo lattice $CeCo_2Ga_8$ using optical pump-probe spectroscopy. Time-resolved pump-probe reflectivity measurements reveal a strong anisotropy in the photoinduced response, which is a direct consequence of the material's unique electronic structure. The temperature dependence of the relaxation dynamics provides evidence for the formation of two distinct hybridization gaps that appear at different temperatures in the heavy fermion state. A direct gap of $2\Delta_{dir} \approx$ 50 meV that persists up to $T^{\dagger} \approx$ 90 K, well above the coherence temperature $T^* \approx$ 20 K. We attribute this higher-temperature gap to the hybridization fluctuations. An indirect gap of $2\Delta_{ind} \approx$ 10 meV opens closer to $T^*$, signifying the development of long-range coherence in the heavy fermion state. Furthermore, we find that the hybridization gap can be suppressed with increasing pump fluence, indicating a delicate interplay between photoexcitation and the coherent heavy fermion state. Our results provide insights into the interplay of Kondo physics and low dimensionality in $CeCo_2Ga_8$, and establish ultrafast optical spectroscopy as a sensitive probe of anisotropic hybridization in heavy fermion materials.

The intricate coupling of spin, orbital, charge, and lattice degrees of freedom in strongly correlated electron systems gives rise to a rich tapestry of emergent phenomena and complex phase diagrams. Heavy fermion (HF) materials, characterized by partially filled 4$f$ or 5$f$ orbitals, exemplify this complexity. In these systems, the delicate balance between the Kondo effect and the Ruderman-Kittel-Kasuya-Yosida (RKKY) interaction governs the low-energy physics, leading to a variety of fascinating ground states, including unconventional superconductivity, quantum criticality, and non-Fermi liquid behavior [1, 2].

A hallmark of HF behavior is the emergence of a hybridization gap, a narrow energy scale arising from the coupling between localized $f$-electrons and conduction electrons. This hybridization process is crucial for understanding the unique properties of HF systems. Traditionally, the nature of $f$-electrons in HF compounds was believed to be temperature-dependent [3, 4]: above the coherence temperature ($T^*$), $f$-electrons were considered localized, while below $T^*$, they were thought to hybridize with conduction electrons, forming delocalized, heavy quasiparticles. Significant progress has been made in recent years, both experimentally, such as angle-resolved photoemission spectroscopy (ARPES) [5–14] and ultrafast optical spectroscopy measurements [15–17], and theoretically [18, 19]. ARPES studies on Ce- [5–10], Yb- [11, 12], and U-based [13, 14] HF systems have challenged this conventional view. These studies reveal the presence of hybridization even at temperatures significantly above $T^*$, suggesting a more nuanced picture of $f$-electron behavior. Furthermore, some ARPES investigations have even suggested the relocalization of $f$-electrons at lower temperatures [8–10, 14], further complicating the understanding of these systems. Ultrafast optical spectroscopy experiments have provided complementary insights, revealing hybridization fluctuations in these materials before the establishment of heavy electron coherence. This results in the formation of an indirect hybridization gap at $T^*$ [15–17]. These intriguing observations highlight the complex nature of $f$-electron behavior in HF materials and underscore the need for further investigation. While significant progress has been made in understanding hybridization dynamics in two- and three-dimensional HF systems, the role of dimensionality remains an open question. It is believed that the nature of unconventional superconductivity, quantum criticality, and non-Fermi liquid behavior are strongly influenced by the dimensionality of the system. Therefore, exploring the effect of dimensionality on hybridization and the resulting emergent phenomena is crucial for a complete understanding of HF systems.

$CeCo_2Ga_8$, a recently synthesized quasi-one-dimensional (quasi-1D) Kondo lattice compound, presents a unique opportunity to explore the interplay of dimensionality and HF physics. This material exhibits a quantum critical point at ambient pressure [20], displaying significant anisotropy in its magnetic susceptibility [20], resistivity [21], and optical spectra [22]. The crystal structure, with Ce atoms forming chains along the $c$-axis isolated by $CoGa_9$ cages [20, 23], leads to quasi-1D heavy electron Fermi surfaces, as predicted by first-principles calculations [20]. Transport measurements reveal incoherent Kondo scattering with an onset temperature of $T_K^{on}$ $\sim$ 90-100 K [21, 24, 25] and the emergence of Kondo coherence at $T^* \approx$ 20 K [21, 24]. Notably, coherence is observed only when the current is applied parallel to the Ce chains [21, 24]. Furthermore, non-Fermi liquid behavior with a linear temperature dependence is observed below 2 K, persisting down to 0.07 K without long-range magnetic order [20, 26]. Recent planar Hall effect (PHE) measurements fur-ther support the presence of hybridization dynamics in this material, indicating an intermediate regime between $T^*$ and $T_K^{on}$, where hybridization remains short-ranged and noncollective [25]. These findings establish CeCo$_2$Ga$_8$ as a rare example of a quasi-1D Kondo lattice in the vicinity of a quantum critical point, offering a platform to investigate the role of reduced dimensionality in driving quantum criticality.

Ultrafast optical spectroscopy offers a powerful lens for investigating the intricate dynamics of low-energy electrons in correlated materials, including high-temperature superconductors [27–31] and HF systems [15–17, 32, 33]. By perturbing these materials with ultrashort laser pulses, time-resolved measurements can effectively disentangle different degrees of freedom based on their characteristic relaxation timescales. This technique is particularly sensitive to subtle changes in the low-energy electronic structure near the Fermi energy ($E_F$), such as the opening of a narrow energy gap. Notably, it enables the study of slow relaxation processes associated with the bottleneck effect in HFs, providing crucial insights into the dynamics of hybridization [15–17, 32, 33].

In this paper, we report an ultrafast optical spectroscopy investigation of the CeCo$_2$Ga$_8$ single crystal, revealing a strong anisotropy in the transient reflectivity that is consistent with its quasi-1D structure. Analysis of the temperature-dependent relaxation dynamics indicates the presence of two distinct hybridization gaps. The first, a direct gap of $2\Delta_{dir} \approx 50$ meV, remains up to $T^\dagger \approx 90$ K, well above the coherence temperature $T^* \approx 20$ K. This higher temperature gap is attributed to the onset of hybridization fluctuations. A second, indirect gap of $2\Delta_{ind} \approx 10$ meV emerges closer to $T^*$, signifying the development of long-range coherence in the HF state.

High-quality single crystals of CeCo$_2$Ga$_8$ were grown using a Ga self-flux method [20]. Ultrafast pump-probe differential reflectivity ($\Delta R/R$) measurements were performed using a 1 MHz Yb-based femtosecond laser oscillator with a center wavelength of 800 nm ($\sim 1.55$ eV) and a pulse width of $\sim 35$ fs. The pump and probe beam polarizations were maintained orthogonal to each other. For a probe beam perpendicular to the $c$-axis of the sample, it is referred to as $s$-polarization, while for a parallel orientation, it is referred to as $p$-polarization. To minimize surface contamination, all measurements were conducted on freshly cleaved surfaces under high vacuum (better than $10^{-6}$ mbar). Further details of the experimental setup can be found in Ref. [27].

Figures 1(a) and 1(b) illustrate the transient reflectivity ($\Delta R/R$) as a function of time delay at various temperatures, measured with $s$-polarization and $p$-polarization, respectively. Upon photoexcitation, we observe a prompt change in reflectivity for both orientations: a decrease for $s$-polarization and an increase for $p$-polarization. This initial response is followed by relaxation processes, gradually returning the system to equilibrium. The relaxation is governed by intricate interactions such as electron-electron and electron-phonon scattering. The photoinduced transient reflectivity exhibits a pronounced temperature dependence for both polarizations. For $s$-polarization, the magnitude of the initial rapid decay component increases significantly with decreasing temperature. In contrast, for $p$-polarization, the initial positive transient reflectivity is enhanced with decreasing temperature. Furthermore, below approximately 90 K, a fast component with a negative amplitude becomes increasingly prominent as the temperature decreases. This low-temperature component exhibits a slower relaxation time and grows in amplitude with decreasing temperature, potentially signaling the opening of a low-energy gap in the density of states [17]. Figures 1(c) and 1(d) show the 2D pseudocolor map of $\Delta R/R$ as a function of temperature and delay time, corresponding to Figs. 1(a) and 1(b), respectively. These plots provide a clearer visualization of the overall trends in the data, making it easier to observe the changes in amplitude and relaxation time. These observations highlight a complex interplay between temperature, probe polarization, and the underlying electronic structure, likely involving the material's quasi-1D Kondo lattice and the hybridization gap.

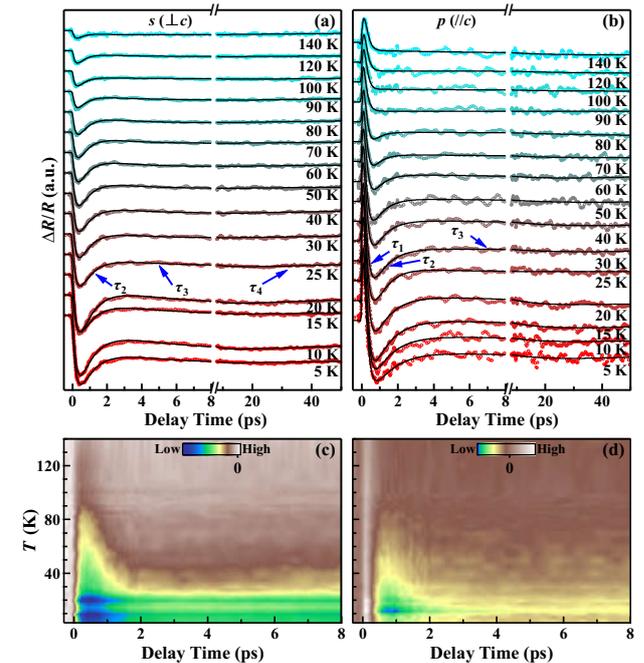

FIG. 1. Photoinduced reflectivity transients ($\Delta R/R$) as a function of time delay at various temperatures for (a) $s$-polarization and (b) $p$-polarization at a pump fluence of 40 $\mu$J/cm$^2$. The black solid lines are Eq. (1) fits. (c) and (d) Corresponding 2D pseudocolor maps of $\Delta R/R$. Relaxation processes: ($i = 1$, $p$-polarization) electron-electron scattering; ($i$=2, $s$ / $p$ ) phonon-assisted electron-hole recombination; ($i = 3$, $s$ / $p$) quasiparticle coupling to bosonic excitations.

To quantify the temperature-dependent behavior, we analyze the quasiparticle dynamics. The solid black lines in Figs. 1(a) and 1(b) represent fits to the transient reflectivity using the following function (See Supplemental Material [34]):

$$\frac{\Delta R(t)}{R} = H(\sigma, t)[\sum_{i=1}^{4} A_i \exp(-\frac{t-t_0}{\tau_i}) + C] \quad (1)$$

where $H(\sigma,t)$ represents the Heaviside step function with an effective rise time $\sigma$. $A_i$ and $\tau_i$ are the amplitude and relaxation time of the $i$th decay process, respectively, and

$C$ is a constant offset. To ensure consistency in component assignments between the *s*- and *p*-polarized data, we performed a four-exponential fit for the *s*-polarized photoinduced transient reflectivity, with $A_1$ constrained to zero. This results in an effective three-exponential decay description across the measured temperature range [Fig. 1(a)]. For *p*-polarization, the relaxation dynamics exhibit a temperature-dependent crossover. Below 90 K, a three-exponential decay function is required for accurate fitting, whereas above 90 K, a two-exponential function (with $A_2 = 0$) suffices [Fig. 1(b)](See Figs. S1 and S2 in the Supplemental Material [34]).

Figures 2(a1) and 2(a2) show the temperature dependence of the amplitude ($A_2$) and relaxation time ($\tau_2$) associated with the fast decay component for *s*-polarization, while Figs. 2(b1) and 2(b2) present the corresponding temperature dependence of $A_2$ and $\tau_2$ for the second decay component observed with *p*-polarization. For *s*-polarization, both $A_2$ and $\tau_2$ increase rapidly with decreasing temperature below $T^\dagger$, showing a notable slope change rather than a sharp transition due to fitting uncertainties. A significant anomaly around 30 K [Figs. 2(a1) and 2(a2)] suggests a possible transition near this temperature. For *p*-polarization, the amplitude and relaxation time parameters show an anomaly around $T^\dagger \sim 90$ K, coinciding with the onset of incoherent Kondo scattering between conduction electrons and localized *f*-moments [20, 25]. Below $T^\dagger$, $A_2$ increases rapidly and continues to grow with decreasing temperature, without any obvious anomalies around 30 K. This contrast highlights a significant anisotropy in quasiparticle relaxation. Similar intermediate-temperature behavior has been reported in other strongly correlated materials, such as heavy fermions [15–17, 35] and high-temperature superconductors [27, 28], often attributed to the formation of hybridization gaps and pseudogaps, respectively. Notably, for *p*-polarization, $A_2$ reaches a maximum slightly below 20 K before decreasing at lower temperatures, possibly due to electron redistribution caused by anisotropic hybridization, warranting further investigation.

The temperature dependence of the amplitude and relaxation time below $T^\dagger$ can be described by the Rothwarf-Taylor (RT) model [36], which captures the time evolution of coupled quasiparticles and bosons in the presence of a narrow energy gap ($\Delta$) in the electron density of states (DOS) near $E_F$. In this framework, excited quasiparticles with energy exceeding the gap recombine, releasing high-frequency bosons that can subsequently re-excite electron-hole pairs. The RT model has been successfully applied to various correlated systems [15–17, 27–31]. According to the RT model, the gap function can be quantitatively extracted from the following equations [31, 32]:

$$\tau^{-1}(T) \propto [D(n_T+1)^{-1} + 2n_T] \quad (2)$$

$$n_T(T) = \frac{A(0)}{A(T)} - 1 \propto (T)^p e^{-\Delta/T} \quad (3)$$

where $n_T$ is the thermal quasiparticle density. The parameter $D$ is a constant that depends exclusively only on the photoex-citation intensity [32]. The parameter $p$ ($0 < p < 1$) reflects the gapped DOS near the gap edge.

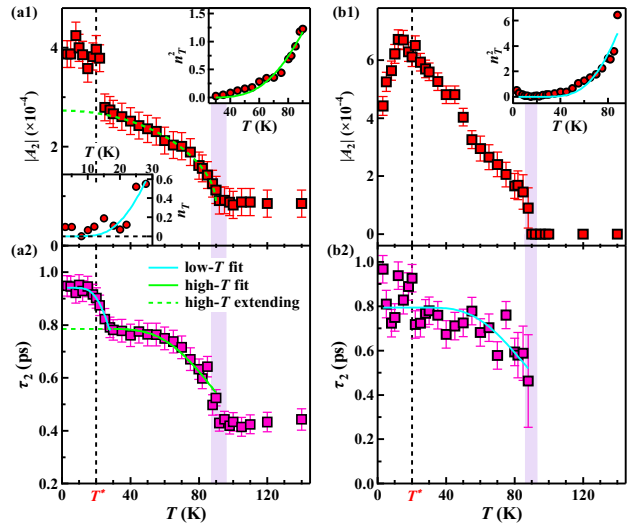

FIG. 2. (a1) Amplitude $A_2$ and (a2) relaxation time $\tau_2$ for *s*-polarization. (b1) Amplitude $A_2$ and (b2) relaxation time $\tau_2$ for *p*-polarization. Insets in (a1) and (b1): calculated $n_T$. Solid lines are fits to the data using the RT model. The green dashed line in (a1) shows a fit to the amplitude (30-90 K) using an empirical mean-field-like temperature dependence, $A(T) = A_2(0)(1-T/T_c)^{2\eta}$, where $T_c = 90$ K and $\eta$ are the characteristic temperature and critical exponent, respectively, to determine $A_2(0)$ for $n_T$ ($T \geq 30$ K) calculations.

Due to the apparent anomaly around 30 K in *s*-polarization, we fit the high- and low-temperature data with the RT model separately. Fitting $\tau_2(T)$ and $n_T^2(T)$ above and below 30 K, using a temperature-independent constant gap $\Delta$, yields energy gaps of $2\Delta_{dir} \approx 50$ meV and $2\Delta_{ind} \approx 14$ meV, respectively, with $p = 0.5$ from a typical BCS-like DOS [32]. We propose that the smaller energy gap, $\Delta_{ind}$, is an indirect hybridization gap emerging slightly above the coherence temperature $T^*$ determined from transport measurements. The larger energy gap, $\Delta_{dir}$, is a direct hybridization gap that persists up to 90 K due to hybridization fluctuations, a phenomenon observed by ARPES in other heavy fermion compounds well above $T^*$ and preceding the emergence of the fully opened indirect hybridization gap at $T^*$, characterized by band bending [5–9]. As indicated by the solid cyan lines in Figs. 2(b1) and 2(b2), fitting $\tau_2(T)$ and $n_T^2(T)$ using a temperature-independent constant gap $\Delta$ yields an energy gap of $2\Delta_{dir} \approx 52$ meV, also with $p = 0.5$. The observed difference in relaxation dynamics between *s*- and *p*- polarizations likely stems from the anisotropic electronic structure caused by its quasi-1D crystal structure. This leads to anisotropic quasiparticle dynamics and uniaxial hybridization in CeCo$_2$Ga$_8$, which is supported by infrared spectroscopy results that hybridization occurs predominantly along the *c*-axis [22].

Although ultrafast spectroscopy lacks the momentum resolution of ARPES, it can still reflect the anisotropy of quasiparticle relaxation, as demonstrated in iron-based high-temperature superconductors with nematic fluctuations [28]. The reflectivity signal contains information about quasiparti-cle relaxation in different momentum space directions. For *s*-polarization, the reflectivity signal is more sensitive to relaxation along the *c*-axis. Conversely, for *p*-polarization, relaxation perpendicular to the *c*-axis contributes more significantly. Therefore, regardless of the polarization direction, the reflectivity can reflect hybridization information and provide the onset temperature $T^\dagger$ of hybridization fluctuations. However, for *p*-polarization, the reflectivity signal is less sensitive to quasiparticle dynamics related to *c*-axis hybridization, explaining the absence of a clear anomaly around 30 K.

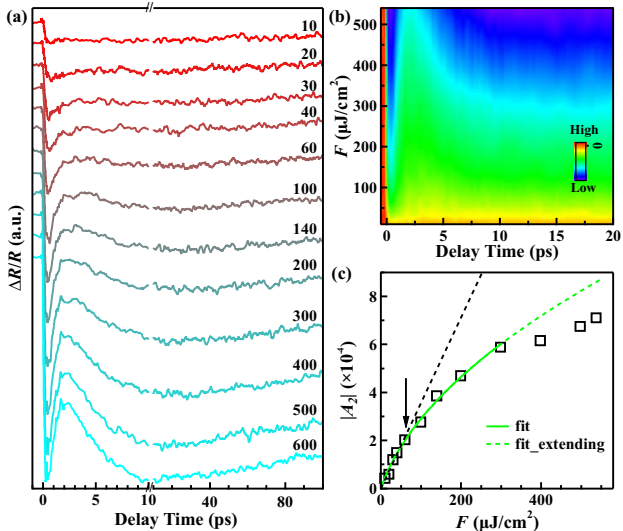

FIG. 3. (a) Transient reflectivity ($\Delta R/R$) as a function of pump-probe delay time for various pump fluences at 5 K (*s*-polarization). Note the break in the *x*-axis. (b) Two-dimensional pseudocolor map of $\Delta R/R$ as a function of pump fluence and delay time at 5 K. (c) Magnitude of the initial decay component amplitude $|A_2|$ as a function of pump fluence. The green solid line represents a fit to Eq. (4), while the black dashed line shows a linear fit to the low-fluence data. The arrows indicate the deviation from the $|A_2| \propto F$ dependence.

To further investigate the underlying mechanism of hybridization and its dependence on excitation density, we performed pump fluence-dependence transient reflectivity experiments for *s*-polarization, as shown in Fig. 3(a). The $\Delta R/R$ signal exhibits a complex dependence on pump fluence, with both the amplitude and the relaxation dynamics changing significantly. Figure 3(b) presents a 2D pseudocolor map of $\Delta R/R$ as a function of pump-probe delay time and fluence, clearly illustrating this fluence dependence. Notably, the evolution of the transient reflectivity with increasing fluence differs significantly from that observed with increasing temperature [Fig. 1(a)], highlighting the distinct nature of the two excitation methods. While increasing temperature results in a decrease in the initial reflectivity change, increasing pump fluence results in a more pronounced initial reflectivity drop accompanied by increasingly strong secondary rising peak. This suggests that the underlying mechanisms at play are different.

The transient reflectivity can be fitted with a three-exponential decay function across the entire measured fluence range, allowing us to extract the fluence dependence of the amplitudes. Figure 3(c) summarizes the fluence dependence of the amplitude $|A_2|$. At low fluence, $|A_2|$ grows monotonically with increasing fluence, exhibiting a near-linear dependence up to $F \approx 60$ $\mu$J/cm$^2$, as depicted by the black dashed line. This initial linear regime suggests a direct relationship between the number of excited quasiparticles and the observed reflectivity change. For HF compounds at low temperatures, the relationship between the amplitude $A$ and pump fluence $F$ has been proposed as [33]:

$$A \propto \sqrt{1 + cF} - 1 \qquad (4)$$

where $c$ is a fitting parameter. We find that Eq. (4) accurately describes the fluence dependence of $|A_2|$ for fluences $F \leq 300$ $\mu$J/cm$^2$, as shown by the solid green line in 3(c). This agreement suggests that, in this lower fluence regime, the ultrafast dynamics are dominated by the physics of the Kondo effect. As the fluence increases, the observed faster relaxation, indicated by the narrowing of the initial negative peak, can be attributed to the gradual suppression of the hybridization gap[33]. As the fluence increases, a larger number of quasiparticles are excited, effectively screening the local moments and reducing the hybridization strength, thus leading to a faster decay. This characteristic fluence of 300 $\mu$J/cm$^2$, where we observe a deviation from the prediction of the RT model, can be compared to similar studies on other HF systems, although direct comparisons should be made with caution due to varying experimental conditions and definitions of this characteristic fluence. For example, CeAuSb$_2$ has been reported to have a similar characteristic fluence of approximately 200 $\mu$J/cm$^2$ at 3.8 K [17], while CeAgSb$_2$ exhibits a change in behavior at a lower fluence, less than 60 $\mu$J/cm$^2$ at 12 K [35], and YbB$_{12}$ requires a higher fluence, exceeding 0.4 mJ/cm$^2$ [37]. These differences may reflect variations in the strength and robustness of the hybridization gap in these materials.

However, beyond 300 $\mu$J/cm$^2$, the behavior of $|A_2|$ deviates from the prediction of Eq. (4). The rate of increase in $|A_2|$ slows down, while the relaxation dynamics evolve, characterized by a more pronounced initial negative peak in the transient reflectivity signal, followed by a more distinct secondary rising peak. This deviation indicates that the strong photoexcitation in this higher fluence regime leads to a qualitative change in the system's dynamics. We attribute this to the intense pumping effectively disentangling the hybridization between the conduction and 4$f$ bands, resulting in the smearing of the hybridization gap or even a photoinduced phase transition [17, 33], along with the emergence of additional relaxation pathways. This disentangling of the hybridization is consistent with the small indirect hybridization gap ($\Delta_{ind}$) observed at low temperatures, which implies a weak hybridization strength that is susceptible to strong photoexcitation. The observed changes in the transient reflectivity profile, including the more pronounced negative peak and more pronounced secondary rising peak, suggest that these additional relaxation processes contribute significantly to the observed signal. Further investigation, possibly involving modeling of the excited state dynamics, would be needed to fully elucidate the complex processes occurring in this high-fluence regime.

In conclusion, we report an ultrafast optical spectroscopy study of the quasi-one-dimensional Kondo lattice CeCo$_2$Ga$_8$. Our measurements reveal strong anisotropy in the transient reflectivity, reflecting the material's unique structure. We observe two distinct types of hybridization gaps: a direct gap that persists up $T^\dagger \approx 90$ K, associated with hybridization fluctuations, and an indirect gap emerging at a lower temperature slightly above $T^* \approx 20$ K, where coherent Kondo screening develops. Analysis of the temperature- and fluence-dependent relaxation dynamics using the Rothwarf-Taylor model provides quantitative estimates of these gaps, highlighting the interplay of Kondo physics and low dimensionality in this system. Our results establish ultrafast optical spectroscopy as a powerful tool for probing anisotropic hybridization and emergent phenomena in heavy fermion materials.

This work was supported by the National Natural Science Foundation of China (Grant No. 12074436, U22A6005, and U2032204), the National Key Research and Development Program of China (Grant No. 2022YFA1604204), the Science and Technology Innovation Program of Hunan Province (2022RC3068), and the Natural Science Foundation of Changsha (kq2208254).

---

abatake, H. Okamura, and J. Demsar, Phys. Rev. B **103**, 115134 (2021).

---

* Corresponding author: jqmeng@csu.edu.cn